# Amplification of Electrostriction Mechanism of Photoacoustic Conversion in Layered Media

Grigory Knyazev[1,2], Daria Ignatyeva[1,2], Ivan Sopko[1], Vladimir Belotelov[1,2], Oleg Romanov[3]

[1]Faculty of Physics, M.V. Lomonosov Moscow State University , 1 Leninskie Gory, 119991 Moscow, RUSSIA
[2]Russian Quantum Center, 100A Novaya Str, Skolkovo, 143025 Moscow, RUSSIA
[3]Faculty of Physics, Belarusian State University, 4 Nezalezhnasti Ave., 220030 Minsk, BELARUS

E-mail: g_knyazev@mail.ru

**Abstract**

In this work we have performed an analysis of electrostriction mechanism of optical to acoustical energy conversion on the interface of two materials with low optical absorption. We compared this method of conversion with widely used thermal conversion based on thin metal film. It was shown, that the contribution of electrostriction mechanism is significantly lower in case of homogeneous medium. We demonstrated the possibility to amplify the generated acoustic signal by excitation of the guided modes, which energy is localized in a thin (of about 200-400 nm thickness) dielectric layer. Due to high electromagnetic field energy concentration in the structure, local values of intensity increase by more than two order of magnitude in comparison to intensity of incident light, that also allows to increase the amplitude of the pressure correspondingly. Thus, in the project we propose the novel layered dielectric structures, in which electrostriction effect occurs due to excitation of the guided modes.

## Introduction

Photoacoustic effect is widely applied in various fields of science, such as medicine, physics and biology, and is commonly utilized in various devices. This effect is utilized for the photoacoustic spectroscopy [1], optoacoustic tomography, and also for biological tissue studies. Photoacoustic effect is used for the nondestructive testing of such characteristics as sound velocity, stress, strain, etc. Recently, studies of the processes of the acoustic pulses excitation in absorbing micro and nanostructures are of particular interest. The frequency of sound oscillations generated in such structures upon absorption of the energy of short (nano- and picosecond) and ultra-short (femtosecond) laser pulses lies in the range from giga- to terahertz [2, 3]. In paper [4] photoacoustic excitation of short acoustic impulses was used to study magnetophotonic nanostructure $Al_2O_3$-Co-Au-air. The precision of acoustic layer thickness measurement exceeded 35 nm. Recently there has been a lot of work with light modulation by photoacoustically excited sub-terahertz pulses. For example, in [5,6] pump-probe measurement technique was used to study layered structures based on photonic crystal resonator and plasmonic grating.

One of the challenges inhibiting the development and improvement of devices, based on the photoacoustic energy conversion, is efficiency enhancement of light radiation energy transfer to acoustic waves.

There are two major methods of sound generation with electromagnetic radiation: linear and quadratic in amplitude of electromagnetic field. In the first case the excited sound has the same frequency as the electromagnetic wave. However, interaction of the laser radiation with the medium occurs due to effects, which are quadratic nonlinear in field. These effects include: electrostriction, magnetostriction, thermal effect and radiation pressure [7,8]. These mechanisms cause additional mechanical stresses in the medium, which serve as the acoustic waves sources. Therein the most effective and most commonly used is the thermal effect.

In this work we consider two mechanisms of acoustic wave excitation by the short high-intensity optical pulse. The first mechanism is widely used in photoacoustics and is based on thermal interaction of optical radiation with the medium. Thermal excitation of acoustic pulses occurs in the following manner: femto-second laser impulse falls on the photoacoustic transducer–a thin absorber film. Due to high absorption coefficient most of the energy transfers into heat. This heating causes the film to expand and create the initial mechanical strain. Acoustic pulse propagates in the medium, while the optical radiation barely penetrates the absorber layer. If the goal of the experiment is the generation of hypersonic acoustic pulses, it is necessary to use thin film photoacoustic transducers as the ultrasound wavelength is defined by the film thickness. However, in order to achieve maximum efficiency it is necessary for the film to absorb all of the optical radiation. Analysis shows, that the necessary absorption coefficient is about $10^5$ cm$^{-1}$. Such absorption is usually achieved with thin metallic films. The major downside of the metallic transducer is very high reflection coefficient, which can lead to reduction of the photoacoustic energy conversion efficiency. However, the effect of light on the material in the case of a metal structure can be increased due to the excitation of plasmon resonances. [9-13, 14]

The second mechanism of ultrasound excitation is based on the electrostriction in transparent dielectric layers. In order to achieve predominance of this mechanism of photoacoustic ultrasound excitation extremely high intensity levels are required. That is why this mechanism is rarely used in practice. However, we show here, that the layered photonic crystals based structures similar to those considered in [15] allows achieving concentration of optical energy high enough for acoustic impulse generation with transparent transducer.

**Photoacoustic interaction mechanism.**

In this section we consider forces, causing elastic strain in the medium induced by optical radiation. The force density in nonmagnetic isotropic medium $I$ can be written as:

$$f = -\nabla p(p_0, T) - \nabla \varepsilon \, I + \nabla((\rho \frac{\partial \varepsilon}{\partial \rho})_T I) + (\varepsilon - 1)\frac{\partial I}{\partial t}, \quad (1)$$

where $p_0$ is equilibrium pressure in the medium, $T$ is temperature perturbation, $\rho$ and $\varepsilon$ are density and dielectric permittivity of the medium. In Equation (1) the first term describes the forces related to optical heating of the medium. The second and third terms are related to electrostriction forces, while the final term is the Abraham force [8]. This force can be non-negligible only in non-absorbing media characterized by homogeneous light intensity distribution. As our work is dedicated to analysis of the structures with inhomogeneous distribution of optical radiation, the last term of Equation (1) can be dismissed.

Strictly speaking, Equation (1) is also valid for media with low absorbance; however, the analysis shows that with the growth of absorption coefficient in case of homogenous medium the first term starts to prevail over others [7]. Because of that, while studying the structure consisting of the gold film, deposited on the surface of GaAs (see Figure 1a), we shall use the following method. As we consider the optical heating to occur within the time intervals about 1 ps, it is appropriate to use two temperature model. Optical radiation transfers its energy to free electrons in metal, which gradually heats up the lattice:

$$\rho_e C_e \, \partial T_e/\partial t = k_e \Delta T_e + \sigma I - \gamma(T_e - T), \quad (2)$$

$$\rho C \, \partial T/\partial t = \gamma(T_e - T), \quad (3)$$

where $\rho_e, C_e, T_e$ and $k_e$ are density, thermal capacity, temperature and thermal conductivity of electron gas, $\sigma$ is optical radiation attenuation coefficient, $\gamma$ is electron-lattice interaction factor and $\rho, C, T$ are density, thermal capacity and temperature of metal [16]. Equation (3) does not take into consideration the thermal conductivity of metal, as at time intervals less than 1 ns its contribution is marginal.

While describing the acoustic wave propagation in the Au-GaAs structure we shall neglect nonlinear effects, which are usually observed in liquid media [17]. Thus, the acoustic wave propagation can be written as:

$$\rho \frac{\partial^2 u_i}{\partial^2 t} - c_{ijk} \frac{\partial^2 u_l}{\partial x_j \partial x_k} = -c_{ijk} \alpha_{kl}^l \frac{\partial T}{\partial x_j}, \quad (4)$$

where $c_{ijk}$ is elastic tensor and $\alpha_{kl}$ is thermal expansion coefficient tensor for Au film. The photoacoustic elastic waves in similar structures have been studied previously [12, 13]. It is worth noting, that in the metal-substrate type structure it is typical to see pulses sequence, as the metal layer acts as an acoustic resonator.

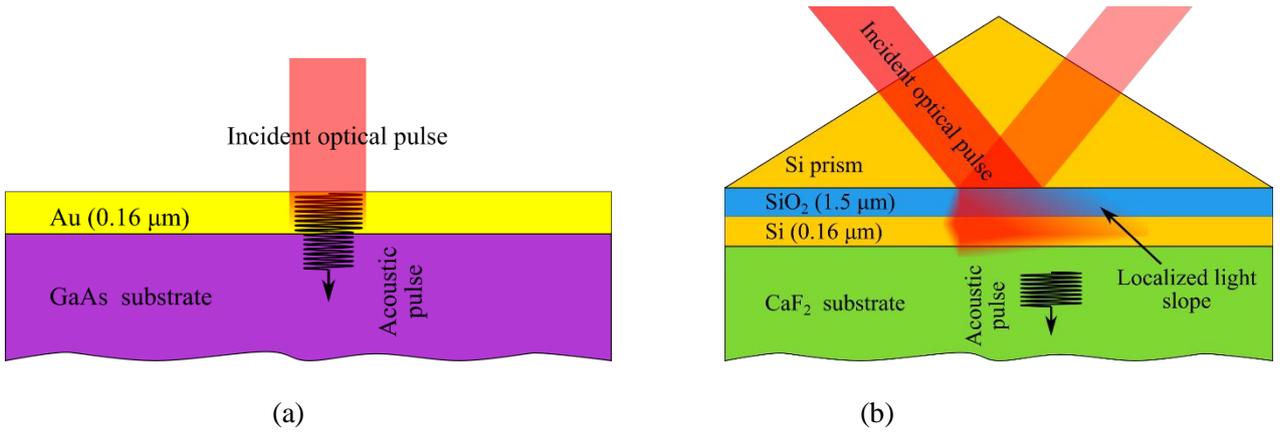

**Figure 1.** Schemes of optoacoustic transducer structures. (a) metallic film-based structure with optical heating mechanism of energy conversion, (b) dielectric structure with electrostriction mechanism of energy conversion

Let's now consider the structure shown in Figure 1b. Optical absorption in this structure at the near infrared is negligibly small. Therefore, the acoustic wave excitation will be mostly described by the second and the third terms of Equation (1). It is worth noting, that in the case of the homogeneous plate the second term is no longer present, $\nabla \varepsilon = 0$. For the layered structure the nonzero second term is present near the edges of the layers. Its contribution is negligible if the thickness of the layer exceeds the value of layer inhomogeneity, which is usually about 10 nm. Thus the Equation (1) can be used to derive wave equation, describing photoacoustic generation and propagation of elastic wave in the medium with low light absorption:

$$\rho \frac{\partial^2 u_i}{\partial^2 t} - c_{ijk} \frac{\partial^2 u_l}{\partial x_j \partial x_k} = -\frac{1}{8\pi} \frac{\partial \sigma_{kli} E_k E_l^j}{\partial x_j}. \quad (5)$$

For the optically isotropic medium the right part of Equation (5) equals to the third term of Equation (1). The tensor $\sigma_{ijk}$ can be expressed through the photoelastic tensor $p_{ijk}$:

$$\sigma_{ijk} = {}^l\varepsilon_{km} p_{mni} \varepsilon_{nl}^j. \quad (6)$$

**Optoacoustic interaction via electrostriction in the structured transparent materials**

If s-polarized light is incident then the condition $k=l$ in Eq.(5) is met. Assuming the absence of gyration in layers of the structure, we can stand that in Eq.(6) $m=n$. Meanwhile, if we limit the considered cases with isotropic media and crystals with cubic or orthorhombic symmetry (the assumption is also valid for some other types of symmetry) [18], we can assign $i=j$ in Equation (6). It is also necessary to consider that for the layered medium the light intensity gradient is normal towards the layers. The case of s-polarized light sets the requirement of $k \neq i$. This reasoning is valid as the substrate is a cubic crystal. Next we will assume that $CaF_2$ substrate is cut normally to [100] axis. We also presume that evaporated layers of silicon and quartz are amorphous polycrystals and thus optically isotropic. Thus the Equation (5) can be expressed as:

$$\frac{\partial^2 u_1}{\partial^2 t} - v_l^2 \frac{\partial S_1}{\partial x} = -\frac{\varepsilon^2 p_{21}}{\rho c} \frac{\partial I}{\partial x}, \quad (7)$$

where $S_1 = \frac{\partial u_1}{\partial x}$ is strain, $I$ is intensity of light, $c$ is velocity of light, $v_l$ is velocity of longitudinal acoustic wave. Equation (7) is based on the assumption that the optical beam width is much greater than the size of the examined area on the structure (the plane wave approximation). This approximation allows to omit the edge effects, which lead to generation of the surface and transverse acoustic waves.

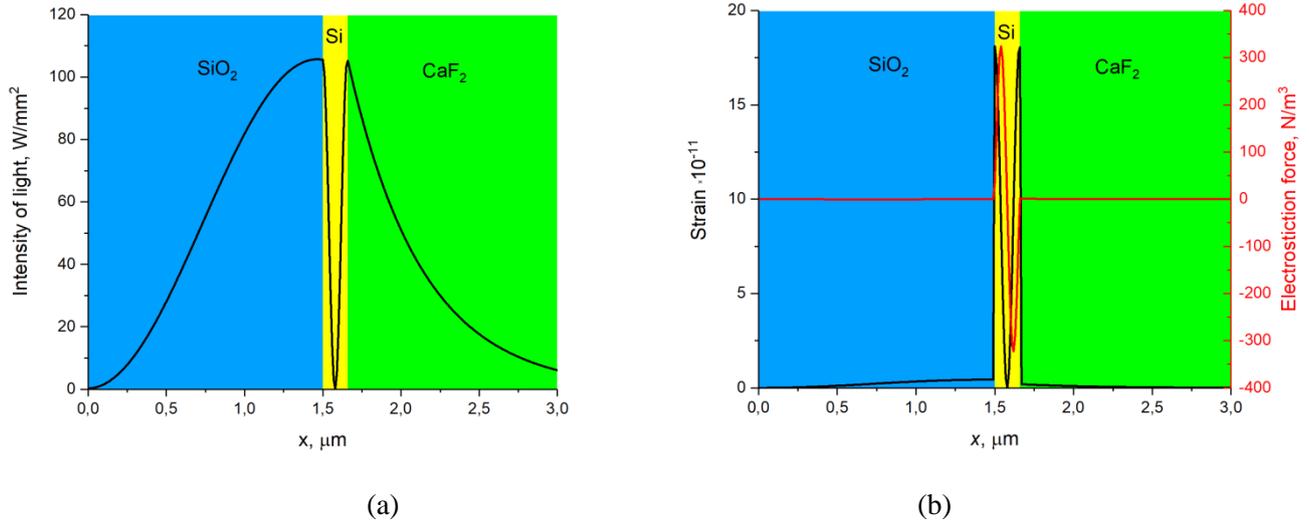

**Figure 2.** Distribution of intensity of light (a) and acoustical strain (b) in the Si-SiO$_2$-Si-CaF$_2$ structure.

In Figure 2 results of modeling of studied dielectric structure are shown. The calculations were performed for the case when the laser pulse is incident on the prism (see Figure 1b) with center wavelength of 1 μm and peak intensity of 1 W mm$^{-2}$. The structure represents a waveguide silicon layer, in which the wave is excited via attenuated total internal reflection. Distribution of optical radiation intensity in Figure 2a demonstrates, that the energy of optical radiation is strongly localized near the waveguide. Due to low thickness of the silicon layer it is possible to achieve rather high radiation intensity gradient in it. Calculation of the initial acoustic deformation in the considered structure (see Figure 2b) shows significant enhancement of the electrostriction effect in the waveguide layer. Moreover, distribution of strains and electrostriction force is close to harmonic, that provides an effective excitation of acoustic pulse with a relatively narrow spectrum. The central wavelength in the spectrum is defined by the thickness of silicon film, wavelength of light and angle of incidence on the structure. It is apparent, that to maximize the excitation efficiency the duration of the optical pulse should be close to half of the acoustic wave period. Considering the parameters of the structure, shown on Figure 1b and sound velocity in silicon of $v_{Si} = 8.4 \times 10^3$ m s$^{-1}$, the duration of the optical pulse should not exceed 10 ps.

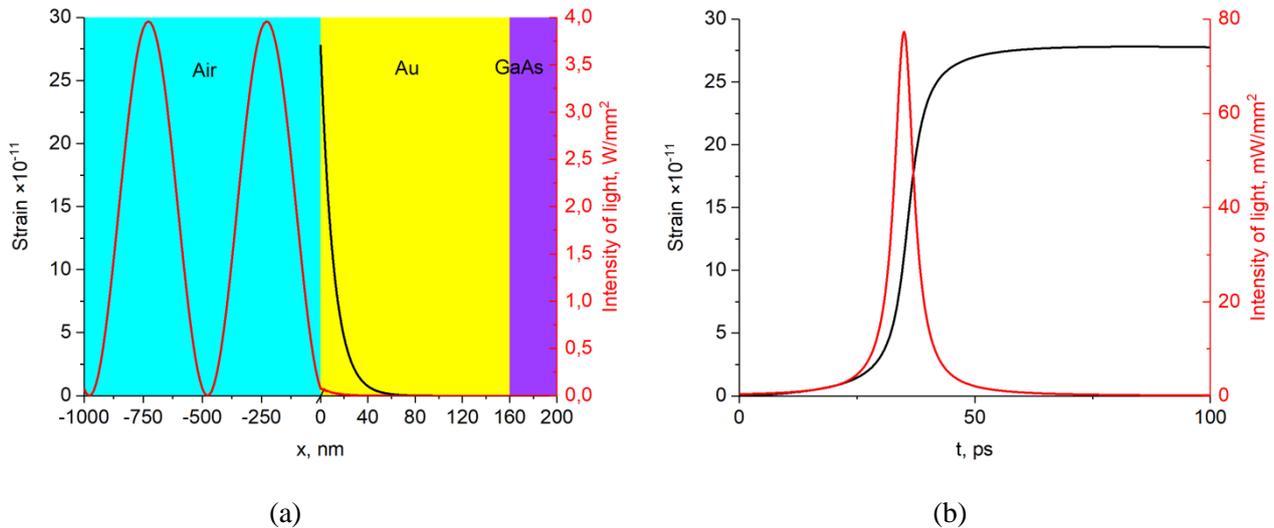

**Figure 3.** Spatial (a) and temporal (b) distribution of light intensity and acoustical strain in the air-Au-GaAs structure.

Thermo-optical generation of the ultrasound in air-Au-GaAs structure is characterized by some other features. We undertook the study of the same parameters of the optical pulse: 1 μm light wavelength, peak intensity of 1 W mm$^{-2}$ and 10 ps pulse duration. It can be seen (see Figure 3a), that the light is predominately reflected from the metal film. This is confirmed by intensity oscillations in air region, caused by the interference of incident and reflected light beams. Meanwhile the intensity of the radiation transmitted into the film is exponentially decaying with the factor of $α=9\ 10^7$ m$^{-1}$ without reaching the GaAs part.

Modeling of thin gold film laser heating using Equations (2), (3) and (4) allows finding the initial distribution of acoustic deformation in the film. It can be seen, that the strain occurs in small confined area of the metal film (see Figure 3a). However, despite of the optical radiation transfer to electrons and then to the lattice, the strain occurs in time interval similar to the laser pulse duration (see Figure 3b). It should be noted, that because of sound reflection from the edges of metal film, acoustic signal penetrates into the GaAs substrate as a series of decaying pulses with pulse repetition interval of $2l/v_{Au}$, where $l$ is metal film thickness [12].

## Discussion

The studied structures show similar level of acoustic strain with discussed above parameters of optical pulse, as shown in Figure 2 and Figure 3. However thermal and electrostriction methods possess several differences and specific features. Application of homogenous metal film for the case of infrared radiation turns out to be inefficient. However, it was shown in several works [4,12,13] that the efficiency can be enhanced by application of the metal grating. In that case the excitation of plasmon resonances can significantly reduce the reflectance. As follows from Equations (2) and (3), with the increase of optical pulse duration thermo-optical energy conversion allows to achieve greater acoustic pulse amplitude, while in case of electrostriction the acoustic wave amplitude does not depend on the laser pulse duration. Consequently, we can state that electrostriction based methods are preferable for subterahertz ultrasound pulses excitation. As shown in Figure 2, the discussed structure represents a one-dimension waveguide. Meanwhile the initial deformation represents the shape of one period of cosine function, which provides a narrower spectrum of acoustic signal in comparison to thermo-optical generation. The narrowest spectrum of acoustic signal can be achieved with multimode waveguide layer. This can be achieved by variation of the thickness of the silicon layer, wavelength of light or angle of incidence. In this case the initial deformation will represent the shape of several period of harmonic function. The center wavelength could be readjusted by tweaking of incidence angle of light on the structure. It should be noted that in case of the air-Au-GaAs structure the shape of acoustical signal is determined by the thickness of the metal film and is fixed for the fabricated structure.

## Conclusion

We have examined the alternative mechanism of optoacoustic energy conversion based on the electrostriction phenomenon in a nonabsorbent medium. Efficiency of this mechanism can be enhanced by application of the layered dielectric structures, supporting excitation of the waveguide modes. In the case of structures based on metal films, the structure undergoes optical heating; therefore, the amplitude of the acoustic wave is determined by the average optical power over the duration of the laser pulse. At the same time, in the case of the electrostrictive method, the amplitude of the acoustic signal is determined by the peak power of the laser pulse. Therefore, with a decrease in the laser pulse duration, the efficiency of the electrostrictive method increases. Thus, the electrostrictive method can be more effective for the excitation of subterahertz acoustic impulses.

## Acknowledgements

This work was financially supported by Russian Foundation for Basic Research (RFBR) (projects 18-52-00042 and 18-29-20113) and Belarusian Republican Foundation for Fundamental Research (BRFFR) (project F18R-141).